
\magnification=1200
\overfullrule=0pt
\baselineskip=20pt
\parskip=0pt
\def\dag{\dagger}

\def\a{\alpha}     
\def\b{\beta}      
\def\g{\gamma}     
\def\d{\delta}     \def\D{\Delta}

\def\p{\pi}        
\def\r{\rho}       
     \def\S{\Sigma}

\def\y{\psi}       

\def\yd{\y^{\dag}}

\def\br{\langle}
\def\ke{\rangle}
\def\ve{\vert}

\def\Winf{$W_{\infty}\  $}
\def\kvecz{\ve z_1 z_2 \cdots z_n\ke}

\def\zbar{\bar{z}}

{\settabs 5 \columns
\+&&&&SU-4240-553 \cr
\+&&&&September 1993\cr}
\bigskip
\centerline{\bf ALGEBRAIC ASPECTS OF THE FRACTIONAL QUANTUM HALL EFFECT }

\vskip 0.4in
\centerline{ Dimitra Karabali {\footnote{*} {e-mail
address: karabali@suhep.phy.syr.edu}}}
\centerline{ Physics Department, Syracuse University, Syracuse, NY 13244}

\vskip 0.8 in
\centerline{\bf Abstract}
\bigskip

Some algebraic issues of the FQHE are presented. First, it is shown that on the
space of Laughlin wavefunctions describing the $\nu =1/m$ FQHE, there is an
underlying $W_{\infty}$ algebra, which plays the role of a spectrum generating
algebra and expresses the symmetry of the ground state. Its generators are
expressed in a second quantized language in terms of fermion and vortex
operators. Second, we present the naturally emerging algebraic structure once a
general two-body interaction is introduced and discuss some of its properties.
\vfill\eject

\noindent{\bf 1. Introduction}

The Quantum Hall Effect $^{[1,2]}$(QHE) appears in two-dimensional
systems of electrons in the
presence of a strong perpendicular uniform magnetic field $B$.
It is characterized
by quantized values of the Hall conductivity, which turns out to be
proportional to the filling factor $\nu$, the ratio between the number of
electrons and the degeneracy of the Landau levels.

The main idea behind the QHE is the existence of a gap, which gives rise to an
incompressible ground state at densities proportional to $\nu$, $\rho=\nu B/2
\pi$. This phenomenon is easily understood in the case of integer $\nu$ (IQHE),
where the first $\nu$ Landau levels are completely filled, because of the large
energy gap between adjacent Landau levels. As a result, the IQHE can be
understood solely in terms of noninteracting fermions. The existence of a gap
is less
obvious in the case of noninteger $\nu$ (FQHE), where Landau levels are only
partially filled. In this case, it is believed that the repulsive Coulomb
interactions
among electrons are important in generating strong correlations which
eventually
produce a new ground state with a gap.

In an attempt to explain the FQHE for $\nu = 1/m$, where $m$ is an odd integer,
Laughlin proposed$^{[3]}$ a set of wavefunctions which turn out to be quite
close to the true solutions,
at least numerically,
for a large class of repulsive potentials$^{[4]}$.
Using the connection to the two-component, one-dimensional plasma, Laughlin
showed that the ground state corresponds to an incompressible configuration of
uniform density $\rho = \nu B/2 \pi$.
Based on Laughlin wavefunctions, a hierarchy scheme$^{[4-5]}$ has
been developed to explain the other rational values of $\nu$. Jain has also
proposed$^{[6]}$ a scheme for constructing generalized wavefunctions which
successfully account for the experimentally observed filling fractions.

  In the limit of very strong magnetic fields we consider all the electrons
confined in the lowest Landau level (LLL), since the energy separation
of adjacent
Landau levels is proportional to the magnetic field.  In ref.[7] we have
presented a
field theoretic formulation for electrons in the LLL and
emphasized the existence of a \Winf algebra, which emerged as the algebra of
unitary transformations preserving the LLL condition and
the particle number. In the thermodynamic limit, in the case of the completely
filled LLL ($\nu =1$) and in the absence of interactions the system is
invariant under \Winf transformations. In the presence of a confining potential
the symmetry is reduced. Some of the \Winf generators annihilate the ground
state while the others create excitations, which include the low lying gapless
edge excitations$^{[8-9]}$. Because the confining potential is itself a member
of the algebra,
\Winf plays the role of a spectrum generating algebra and expresses
the symmetry of the ground state. This point has also been discussed
extensively in ref.[10].

In this paper we first generalize these ideas to the $\nu =1/m$ FQHE as
expressed in terms of the Laughlin wavefunctions. Because of the specific form
of these wavefunctions we find the existence of an isomorphic algebra which
captures the symmetry of the Laughlin ground state.{\footnote{*} {General
arguments hinting at the existence of some infinite dimensional algebra
underlying
the $\nu =1/m$ Laughlin wavefunctions were mentioned
in ref.[10].}} The generators of this algebra are
expressed in a second quantized language and provide a one-parameter family of
\Winf representations, the parameter being related to the filling fraction
$\nu$.

Going beyond the Laughlin wavefunctions and the hierarchy scheme associated
with them, one would like to understand the underlying dynamics responsible
for the FQHE and
the nature of the true ground state. It is essential then to include in the
analysis of the problem the Coulomb interactions among the electrons. Following
the spirit of our earlier algebraic treatment of the IQHE we propose to study
the new algebraic structure emerging
from the introduction of the two-body interactions projected onto the lowest
Landau level. We find that \Winf is naturally extended to a new
infinite algebra, which we refer to as the $X_{\infty}$ algebra, which
contains \Winf as a
subalgebra.

This paper is organized as follows. In section 2 we give a brief discussion of
the field theory of fermions in the lowest Landau level and introduce the \Winf
algebra as the algebra of unitary transformations which preserve the LLL
condition and the particle number. In section 3 we discuss the
role of \Winf algebra in the presence of a confining potential. In section 4 we
project the Coulomb potential onto the lowest Landau level and identify the
Haldane potential whose zero energy eigenstates are the Laughlin states. We
also present a second quantized expression for the vortex operator and
following Read's idea$^{[11]}$ we construct an operator that generates the
Laughlin
ground state from the Fock vacuum. In section 5 we present the infinite
dimensional algebra
which expresses the symmetry of the Laughlin ground state. This is isomorphic
to \Winf . Its generators are given in terms of fermionic and vortex operators.
In section 6 we analyze the enhanced infinite algebraic structure once two-body
interactions are introduced and we conclude with discussions in section 7.
\bigskip
\noindent{\bf 2. Field Theory of Fermions in the Lowest Landau Level}

The many-body Hamiltonian of a system of massive
(mass $M$) fermions in a uniform magnetic
field in two space dimensions is given by
$$\eqalignno{
H_{0}&={1\over{2M}}\sum_{a=1} ^N ( {\bf \Pi }_a )^2
\cr &={1\over{2M}}\sum_{a=1} ^N ( \Pi^x _a +i\Pi^y _a )
( \Pi^x _a -i\Pi^y _a )+ {B\over{2M}}N , &(2.1)\cr}
$$
where
$$
\Pi^i =  p_i -A^i ({\bf x}) =-i {\partial \over \partial x^{i}}-A^i ({\bf x}),
\ \ \ \ \  \ \ \ \  i= x, y \eqno (2.2)
$$
and $B$ is a uniform external magnetic field defined by $\vec{\nabla} \times
\vec {A}=-B$. In what follows we shall omit the last constant term in (2.1).
The corresponding Schr\"odinger wave function
$\Psi ( {\bf x}_1 , {\bf x}_2 , \cdots {\bf x}_N
)$ is a totally antisymmetric function of
${\bf x}$'s.
As is well known, the spectrum consists of infinitely degenerate levels of
energy $E_{n}=\omega n$, where $\omega = B/M$ is the cyclotron
frequency (in units $\hbar=c=e=1$). These are the Landau levels.
For large $B$ or small $M$ the energy
gap between Landau levels is big and to a good approximation we can consider
the fermions restricted to the lowest Landau level. In this case the
fermionic wavefunction obeys $$
( \Pi^x _a -i\Pi^y _a ) \Psi ( {\bf x}_1 , {\bf x}_2 ,
\cdots {\bf x}_N )=0\eqno (2.3)
$$
In the symmetric gauge, where $A^{x}= By/2$, $A^{y}= -Bx/2$, the LLL
condition is written as
$$
(\partial _{z_{a}} + {\textstyle {1 \over 2}}\bar{z}_{a} ) \Psi ({\bf x}_1
, {\bf x}_2 , \cdots {\bf x}_N )=0
\eqno(2.4)
$$
where $z_{a} = \sqrt{{B \over 2}}(x_{a}+iy_{a})$,
$\bar{z}_{a} = \sqrt{{B \over 2}}(x_{a}-iy_{a})$. The solution to (2.4) is
given
by
$$ \Psi ( {\bf x}_1 , {\bf x}_2 , \cdots {\bf x}_N
 ) = f(\bar{z}) \exp ({-{\textstyle {1 \over 2}}\sum _{a=1}
^{N} |z_{a}| ^2}) \eqno (2.5)
$$
where $f(\bar{z})$ is a polynomial in the variables $\bar{z}_{a}$.

The above many-body  quantum mechanical system given in (2.3)-(2.5) can be
further expressed in a second quantized
language. The fermion operator satisfies the constraint
$$
\left( \partial_z + {\textstyle {1 \over 2}} \bar{z}\right) \Psi (x,y,t)=0
\eqno (2.6) $$
which again produces the LLL condition (2.4).
A general solution of (2.6) has the form
$$
\eqalign{
\Psi (x,y,t) & = {\sqrt{B\over {2\pi}}}\ e^{-{1\over 2}|z|^2 }
\sum_{n=0} ^{\infty}{{\bar{z} ^n} \over {{\sqrt {n!}}}}
C _n (t) \cr
& \equiv  {\sqrt{B\over {2\pi}}}\ e^{-{1\over 2}|z|^2 }  \psi (\bar{z} ,t) \cr}
\eqno (2.7)
$$
where the modes $C_n$ satisfy the usual anticommutation relations
$\{C_n, C^{\dag} _m
\}=\delta _{nm}$. The constrained  fermion operators no longer satisfy the
usual anticommutation relations. One can show that
$$
\{ \psi (\bar{z},t), \psi^{\dag} (z',t) \}= e^{z'\bar{z}} \eqno (2.8)
$$
For later purposes let us define $|n>$ and $|z>$ to be the number basis and
coherent basis representations for a harmonic oscillator
$$ a^{\dag} a |n> = n|n> ~~~~ a|z> = z|z>~~~~ [a,a^{\dag}]=1
\eqno (2.9) $$
The right hand side of eq.(2.8) is essentially a $\delta$-function in a
coherent state representation.

In the absence of an external potential and interactions the maximal symmetry
the system can possess is the \Winf  symmetry$^{[7,10]}$.
The \Winf transformation is defined as a unitary transformation
in the space of ${C}_n$'s:
$$
{C}_n (t)= u_{nm} {C}_m (t)=\br{n} \ve u \ve m \ke {C}_m (t) \eqno
(2.10)
$$
An infinitesimal unitary transformation
is generated by a hermitian operator which we write as
$\ddag \xi ( \hat {a} , {{\hat{a}}^{\dag}} )\ddag$ with the antinormal ordering
symbol, where $\xi$ is a real function when $\hat{a}$ and ${\hat{a}}^{\dag}$
are replaced by $
z$ and $\bar z$ respectively. Then using (2.7) we obtain the following
infinitesimal transformation for ${\Psi}$:
$$
\delta {\Psi} (x,y,t) =i{\sqrt{B\over {2\pi}}}\ e^{-{1\over 2}|z|^2 }
\ddag \xi ( \partial_{\bar{z}} , \bar{z}  )\ddag \sum_n \langle{z}\ve n\ke
{C}_n (t)
=i\ddag \xi ( \partial_{\bar{z}}+{1\over 2}z , \bar{z}  )\ddag {\Psi} (x,y ,t)
\eqno (2.11)
$$
where $\ddag\ \ \ \ddag$ indicates that the derivatives are
placed on the left of $z$ and $\bar {z}$. The fermion density $\rho =  \Psi
^{\dag} \Psi $ transforms as
$$
\delta \rho (x,y,t) = i (\ddag \xi (\partial_{\zbar} +z, \zbar) \ddag - \ddag
\xi (z, \partial_z + \zbar)\ddag) \rho (x,y,t)
\eqno (2.12) $$
while the fermion number remains invariant
$$ \int dx dy \delta \rho (x,y,t) =0 \eqno (2.13) $$
The generator of this \Winf transformation is given by
$$
\rho [{\xi}] \equiv \int dxdy \xi (z, \bar{z}) \rho (x, y, t)
=\int  d^2 z e^{-|z|^2} \psi ^{\dag} (z)
\ddag\xi ( \partial_{\bar{z}} , \bar{z} )\ddag
\psi (\bar{z}) \eqno (2.14) $$
where $d^2 z \equiv {B \over {2 \pi}} dx dy$.
The operators $\rho[\xi]$ satisfy an infinite
dimensional algebra given by
$$
[\,\rho [\,\xi_1 \, ],\rho [\, \xi_2 \, ]\, ]={i\over B}\rho
[\{\!\!\{\xi_1 ,\xi_2 \}\!\!\}] \eqno (2.15)
$$
where
$$
\{\!\!\{\xi_1 ,\xi_2 \}\!\!\}=iB{\sum_{n=1} ^{\infty}}{{(-)^n}\over{n!}}
\left(
{\partial_{z} ^{n}}\xi_1 {\partial_{\bar{z}} ^n}\xi_2 -
{\partial_{\bar{z}} ^{n}}\xi_1 {\partial_{z} ^n}\xi_2\right)\eqno (2.16)
$$
$\{\!\!\{ \}\!\!\}$ is the so called Moyal bracket. By choosing
$\xi (z,\bar{z})=z^l \bar{z}^m$ we obtain the commutation relation
$$[\, \rho_{rs} \, , \rho_{lm} \,]= \sum_{n=1}^{min(s,l)} {{(-)^n} \over n!}
{{l!s!} \over {(l-n)!(s-n)!}} \rho_{r+l-n,s+m-n} - (s \leftrightarrow m ,
r \leftrightarrow l) \eqno(2.17)
$$
where $\rho_{lm}= \int d{\bf x} z^l \bar{z}^m \rho ({\bf x}) \equiv \rho [z^l
\bar{z}^m]$. The Lie algebra (2.15) and its representation (2.17) in the
specific basis are manifestations of the $W_{\infty}$ algebra$^{[12]}$,
which in this
case is the algebra of $U(\infty)$. It corresponds to
unitary transformations (linear in the space of $C$'s) which preserve the
lowest
Landau level condition and the particle number.

The field operator $\Psi(x,y,t)$ is expanded, in equation (2.7), in terms of
one-body
angular momentum eigenstates. Since we work in the infinite plane, no
boundaries are present, all different angular momentum eigenstates are
degenerate.
In order to confine the electrons in a finite area one can introduce an
external potential $V_c ({\bf x})$ which splits the degeneracy and
associates higher energy to higher angular momentum states.
In particular, we choose a harmonic oscillator potential whose projection onto
the lowest Landau level takes the form
$$V_c= \lambda \int d^2 z e^{-|z|^2}  \psi^{\dag}(z) \partial _{\bar{z}}
\bar{z} \psi (\bar{z}) \eqno(2.18)$$
where $\lambda$ is a positive constant.

Since the confining potential is a member of the algebra, $V_c=\lambda
 \rho_{11}$, eq.(2.15) plays the role of a spectrum
generating algebra.

\vskip .3 in

\noindent{\bf $W_{\infty}$ algebra and $\nu =1$ QHE}

For the case of the completely filled LLL ($\nu =1$), the presence of
the confining potential selects a unique ground state which is the minimum
angular momentum state
$$ \Psi ^0 _1 ({\bf x_{1}}, {\bf x_2},...,{\bf x_N}) =
\prod_{i<j} (\bar{z}_{i}-\bar{z}_{j}) \exp({-{\textstyle {1 \over 2}}
\sum_{i=1}^N |z_i|^2}) \eqno (3.1)$$
This corresponds to a circular droplet configuration which is
incompressible. Compression of the droplet corresponds to lowering its
angular momentum, which would require an electron to jump to the next Landau
level. However this is not energetically allowed due to the big energy gap.
On the other hand, deformations that would result in transitions to states
with higher angular
momentum are allowed and cost some energy due to the confining potential.
These higher angular momentum states are states of the form$^{[8]}$
$$ \Psi_1 ({\bf x_{1}}, {\bf x_2},...,{\bf x_N}) = \prod_{i<j}
(\bar{z}_{i}-\bar{z}_{j}) P(\bar{z}_1,...,\bar{z}_N)
\exp({-{\textstyle {1 \over 2}}
\sum_{i=1}^N |z_i|^2}
)\eqno (3.2)$$
where $P(\bar{z}_1,...,\bar{z}_N)$ is a symmetric polynomial. If the polynomial
$P$ is homogeneous, the above states are energy eigenstates (given (2.18))
with the energy depending on the degree of the polynomial. In particular
when the degree of the polynomial $P$ is much smaller than $N$, of order O(1),
the states (3.2) correspond to edge excitations$^{[8-9]}$.

It is clear from our discussion in the previous section that a deformation of
the ground state which produces the excited states (3.2) has to be generated by
a \Winf transformation. Using eq.(2.8) we find that
$$
\rho[\xi]|\Psi_1^0>= \int d^2 z_1...d^2 z_N
e^{-\sum_{i} |z_i|^2} \sum_{k}
\ddag\xi(\partial_{\bar{z}_k},\bar{z}_k )\ddag
\prod_{i<j} (\bar{z}_i-\bar{z}_j)|z_1...z_N>  \eqno(3.3) $$
where $|\Psi_1^0>$ is the ground state of the completely filled lowest Landau
level
$$
|\Psi_1^0>= \int d^2 z_1...d^2 z_N e^{-\sum_{i}
|z_i|^2} \prod_{i<j}
(\bar{z}_i-\bar{z}_j) |z_1...z_N>
\eqno(3.4)$$
and $|z_1...z_N>={1 \over
\sqrt{N!}}\psi^{\dag}(z_1)...\psi^{\dag}(z_N)|0>$.
Since in general
$$ \sum_{k} \ddag\xi(\partial_{\bar{z}_k},\bar{z}_k)\ddag \prod_{i<j}
(\bar{z}_i- \bar{z}_j)= \prod_{i<j} (\bar{z}_i-
\bar{z}_j)P(\bar{z}_1,...\bar{z}_N) \eqno(3.5)$$
where $\xi (z,\zbar)$ is a polynomial in $z$, $\zbar$ and $P$ is a symmetric
polynomial, we find that
$$ \eqalignno{
& \rho_{lk} |\Psi_1^0>=0 ~~~~~~~~~~~~  {\rm if}~~l>k \cr
& \rho_{lk} |\Psi_1^0>=|\Psi_1> ~~~~~~    {\rm if}~~l\le k
&(3.6) \cr}
$$
The first line in eq.(3.6) expresses the symmetry of the
ground state and the second line the generation of
excitations$^{[10]}$. In particular the excitations generated
by $\rho_{l,l+i}$, with
$i<<N$ correspond to edge excitations$^{[8-9]}$.

Since the Hamiltonian is a member of \Winf , the excitation spectrum
is also dictated by \Winf
$$[V_c ~ , ~ \rho_{l,l+k}]= \lambda ~k~ \rho_{l,l+k} \eqno(3.7)$$

 Therefore the \Winf algebra serves as the spectrum generating
algebra for the excitations (3.2) and
captures the infinite
symmetry of the ground state (as $N \rightarrow \infty)$.

\bigskip
\noindent{\bf 4. Laughlin States and Vortices}

While the IQHE can be understood in terms of noninteracting fermions, repulsive
Coulomb interactions and the resulting correlations among the electrons are
believed to play a crucial role in the emergence of FQHE.

Since our analysis
has been restricted to LLL electrons, which is believed to capture the
essential physics, we have to project any relevant potential onto the lowest
Landau
level$^{[13]}$.
The projected Coulomb interaction is given by
$$
V_{int} =\int d \vec{x} d \vec{x}' :\r (\vec {x})\r
(\vec{x}'):v(|\vec{x}-\vec{x}'|)
=\int d^2 z d^2 z'e^{-|z|^2}e^{-|z'|^2} \yd (z)\yd (z')\y (\zbar ')\y (\zbar )
v(\sqrt{2\over B}
|z-z'|) \eqno (4.1)
$$
where
$$
v(|\vec{x}-\vec{x}'|)={1 \over{|\vec{x}-\vec{x}'|}}\eqno (4.2)
$$

Next let us define the operator
$$
O_{mn}=\int d^2 z d^2 z'e^{-|z|^2} e^{-|z'|^2} \big({{z-z'} \over {\sqrt{2}}}
\big)^{m} \big({{z+z'}
\over {\sqrt{2}}}\big)^n {1 \over {\sqrt{n!m!}}} \psi(\bar{z})\psi(\bar{z}')
\eqno(4.3)
$$
This is an operator which annihilates a pair of LLL electrons with
relative angular
momentum $m$ (necessarily an odd integer)
and center of mass angular momentum (relative to the origin)
$n$. $\bar{O}_{mn}$ is defined as the hermitian conjugate of $O_{mn}$.
Equation (4.1) can be written in terms of $O, \bar{O}$ as
$$
V_{int} =\sum_{m=odd} v_m\sum_{n=0}^{\infty} \bar{O}_{mn} O_{mn}\eqno (4.4)
$$
where
$$
v_m =\int d^2 z \, e^{-|z|^2} {{|z|^{2m}}\over m!} v({2 \over
\sqrt{B}}|z|)\eqno
(4.5)
$$
The above expressions are general and hold for any two-body interaction. In the
case of the Coulomb interaction the parameters $v$ are monotonically
decreasing,
$v_1 > v_3 > \cdots >0$.

Most of our theoretical understanding of the $\nu=1/m$ FQHE is based on the
Laughlin wavefunctions
$$\Psi_{m}(\vec{x}_{1},...,\vec{x}_{N})= \exp (-{\textstyle {1
\over 2}}\sum_{i} |z_i|^2) \prod_{i<j} (\bar{z}_i-\bar{z}_j)^{m}
P(\bar{z}_1,...,\bar{z}_N)
\eqno(4.6) $$
where $m$ is an odd integer.

Since the Laughlin states involve electron pairs whose relative angular
momentum is larger than or equal to $m$, we obtain
$$
O_{m'n}\ve \Psi_{m} \ke =0\ \ \ {\rm for}\  m' < m\ {\rm and\  all}\ n
\eqno (4.7)
$$
So the space of Laughlin states can be thought of as the space of zero energy
eigenstates of a truncated Coulomb potential
$$V_{int} ^m =\sum_{n={\rm odd}} ^{m-2} v_{n} \sum_{l}\bar{O}_{nl} O_{nl}
\eqno(4.8)
$$
This is the Haldane potential$^{[14]}$.

Here and in the next section we are going to assume that the relevant two-body
interaction is (4.8) and the resulting physical space is the space of the
Laughlin wavefunctions (4.6). We shall show that in this restricted space
there is an underlying \Winf algebra, which plays the same role which the usual
\Winf algebra (2.15) had in the $\nu =1$ case, namely part of it expresses
the infinite symmetry of the Laughlin ground state and part of it generates
excitations.

Before we derive the specific form of the generator of this \Winf algebra, it
is useful to mention some of the properties of the Laughlin states and the
quasihole (vortex) operator. In particular we shall derive a second quantized
expression for the quasihole operator which will emerge later in the
expression of the \Winf generator in the $\nu =1/m$ case.

Let us define the operator
$$\alpha (\bar{z})=\int d^2 z' e^{-|z'|^2} \ln (\bar{z}-\bar{z}')
\psi^{\dag}(z') \psi (\bar{z}') \eqno (4.9) $$
One can show by a straightforward calculation that
$$ \eqalignno {
 \psi(\bar{z}') e^{n \alpha (\bar{z})} & = (\bar{z}\
-\bar{z}')^{n} e^{n \alpha (\bar{z})} \psi(\bar{z}') \cr
e^{n \alpha(\bar{z})}\psi^{\dag} (z')  & = (\bar{z} -
\partial_{z'})^{n} \psi ^{\dag} (z') e^{n \alpha (\bar{z})}
&(4.10) \cr}
$$
where $n$ is an integer.

We first remark that although the operator $\alpha (\bar{z})$ is ill defined
because of a logarithmic singularity, the exponentiated form
$e^{\alpha (\bar{z})}$ is
well defined. The proof of this goes as follows.
 We first observe, using (4.10), that
$$ e^{\alpha (\bar{z})} |z_1...z_N> = \prod _{i=1} ^N (\bar{z}-\partial
_{z_i}) |z_1...z_N> \eqno(4.11)$$
We can then show that
$$e^{ \alpha (\bar{z})} = |0><0| + \sum _{N=1}^{\infty}\int \prod _{i=1}^{N}
(d^2 z_i e^{- |z_i|^2} (\bar{z}-\bar{z}_i))
|z_1...z_N><z_1...z_N| \eqno(4.12) $$
This is true since $|z_1...z_N>$ form a
complete basis and the rhs of the above equation satisfies (4.11). Further
using the fact that the projector on the ground state
can be expressed in terms of a normal product as
$$ |0><0|= : \exp (-\int d^2 z e^{-|z|^2} \psi ^{\dag} (z) \psi (\bar{z})):
\eqno(4.13) $$
it is straightforward to derive that
$$e^{\alpha (\bar{z})} = :\exp (\int d^2 z' e^{-|z'|^2} \psi ^{\dag} (z')
\psi (\bar{z}')[(\bar{z}-\bar{z}')-1]): \eqno(4.14) $$
Expressions (4.12) and (4.14) are free of singularities.

We now claim that $e^{\alpha(\zbar )}$ is the creation operator of
a vortex (quasihole),{\footnote{*} {Our explicit second quantized form of the
vortex operator satisfies the consistent operator equations discussed in
ref.[2] pg. 294.}}
since using (4.11) or (4.12) we have
$$
e^{\alpha (\zbar )}| \Psi_m^0 > = \int d^2 z_1...d^2 z_N
e^{-\sum_{i}|z_i |^2}  \prod_{i<j}(\zbar _i -\zbar _j )^m
\prod_i (\zbar -\zbar _i )\kvecz \eqno (4.15)
$$
where
$$|\Psi _m ^0>= \int d^2 z_1...d^2 z_N
e^{-\sum_{i} |z_i |^2}  \prod_{i<j}(\zbar _i -\zbar _j )^m \kvecz
\eqno (4.16)$$
and $ e^{-{\textstyle{1 \over 2}} \sum_{i} |z _i|^2}
\prod_{i<j}(\zbar _i -\zbar _j )^m
\prod_i (\zbar -\zbar _i )$ is the quasihole wavefunction$^{[3]}$.
In fact using (4.15), (4.16)  we can also show that
$$
\psi (\zbar ) | \Psi _m^0 \rangle _N =\sqrt{N} e^{m \alpha (\zbar )}
 | \Psi _m^0 \rangle _{N-1} \eqno (4.17)
$$
where the subscripts indicate the fermion number.
This verifies that $m$ vortices are equivalent to one hole.

Further, following Read's idea$^{[11]}$ we can construct the operator
$$
q_m^{\dag} =\int d^2 z e^{-|z |^2} \yd (z) e^{m \alpha (\zbar)}
\eqno (4.18)
$$
such that the $N$-body Laughlin ground state is created out of the vacuum as
$$
| \Psi_m^0> = {1\over\sqrt{N!}}(q^{\dag}_{m} )^N \ve 0\ke
 \eqno (4.19)
$$
By using (4.10) and the commutativity of $\alpha$'s, it is not difficult to
prove that
$$
q_m^{\dag} q_m^{\dag} = -(-)^m q_m^{\dag} q_m^{\dag} \eqno (4.20)
$$
Thus for $m$ odd, $q_m^{\dag}$ ($q_m$) is a bosonic operator.

\vskip .3 in
\noindent{\bf 5. $W_{\infty}$ algebra and Laughlin states for $\nu=1/m$}

As we mentioned earlier,
in the presence of the Haldane potential all Laughlin states of the form (3.2)
are degenerate.
The introduction of a confining potential splits the degeneracy of these states
just as in the $\nu =1$ case. The ground state corresponds to the
state with $P=1$, while states with $P \ne 1$ correspond to excited states
of higher energy, including
edge excitations when the degree of the polynomial $P$ is much smaller than
$N$. The states (3.2) form a Hilbert space ${\cal{H}}_m$.

Repeating the steps of section 3, we shall try to construct operators which
generate the states (4.6) by acting on the Laughlin ground state. More
generally we are interested in finding the transformations which preserve the
particle number and
leave ${\cal{H}}_m$ invariant. In doing so we utilize the one-to-one
mapping between $\nu=1$ states (3.2) and $\nu={ 1 \over {2p+1}}$ Laughlin
states (4.6).

Let $|\Psi_{m}^{0}>$ be the $N$-body Laughlin ground state ($m=2p+1$).
Obviously the wanted operator cannot be $\rho[\xi]$ since
$$\rho[\xi]|\Psi_m ^0>=\int d^2 z_1... d^2 z_N e^{-\sum |z_i|^2} \sum_k
\ddag \xi (\partial_{\bar{z}_k}, \bar{z}_k) \ddag \prod_{i<j}
(\bar{z}_i-\bar{z}_j) ^m |z_1...z_N>  \eqno(5.1)
$$
and in general
$$\sum_k
\ddag \xi (\partial_{\bar{z}_k}, \bar{z}_k) \ddag \prod_{i<j}
(\bar{z}_i-\bar{z}_j) ^m \ne \prod_{i<j}
(\bar{z}_i-\bar{z}_j) ^m  P(\bar{z}_1,...,\bar{z}_N) \eqno(5.2) $$
where $\xi (z,\zbar)$ is a polynomial in $z$, $\zbar$ and $P$ is some
symmetric polynomial ($P$ could be also zero).
The operator $\rho[\xi]$ does not keep ${\cal{H}}_m$ invariant since its
action does not preserve the fundamental
property of the
Laughlin wavefunctions to behave like $(\bar{z}_i-\bar{z}_j)^m$.

Let us define the following operators
$$\eqalignno{
U_{2p}&= \sum _{N=2} ^{\infty} \int d^2 z_1...d^2 z_N e^{-\sum |z_i|^2}
\prod_{i<j}
(\bar{z}_i-\bar{z}_j)^{2p} |z_1...z_N><z_1...z_N| \cr
\tilde{U}_{2p} &= \sum _{N=2} ^{\infty} \int d^2 z_1...d^2 z_N e^{-\sum |z_i|^2
} \prod_{i<j}
(\bar{z}_i-\bar{z}_j)^{-2p} |z_1...z_N><z_1...z_N| &(5.3) \cr}
$$
The operator $\tilde{U}_{2p}$ is in general singular but its action on
${\cal{H}}_m$ is well defined. In particular
$$ \eqalignno{
\tilde {U} _{2p} & : ~~~{\cal{H}}_m \rightarrow {\cal{H}}_1 \cr
U_ {2p} & : ~~~ {\cal{H}}_1 \rightarrow {\cal{H}}_m &(5.4) \cr}
$$
Using eqs.(3.3) and (3.5) we find that
$$ \eqalign{
U_{2p}~\rho[\xi] ~\tilde{U}_{2p}|\Psi_m ^0> &= \int d^2 z_1...d^2 z_N e^{-\sum
|z_i|^2}
\prod_{i<j}(\bar{z}_i-\bar{z}_j)^{2p} \sum_{i} \ddag
\xi(\partial_{\bar{z}_i}, \bar{z}_i )\ddag \prod_{k<l}
(\bar{z}_k-\bar{z}_l) |z_1...z_N> \cr
& = \int d^2 z_1... d^2 z_N
e^{-\sum |z_i|^2}
\prod_{i<j} (\bar{z}_i-\bar{z}_j) ^m P(\bar{z}_1,...,\bar{z}_N) |z_1...z_N>
\cr} \eqno(5.5) $$
It is clear now that
$$W_{2p}[\xi] \equiv U_{2p}~\rho[\xi] ~\tilde{U}_{2p} ~~~~~ W_{2p}: {\cal{H}}_m
\rightarrow {\cal{H}}_m \eqno(5.6) $$
is the generator of transformations which preserve the particle number and keep
${\cal{H}}_m$ invariant and also generate excitations in ${\cal{H}}_m$
the same way that
$\rho [\xi]$ generated excitations in ${\cal{H}}_1$.
 In particular
$$ \eqalignno{
& (W_{2p})_{lk} |\Psi_m^0>=0 ~~~~~~~~~~~~  {\rm if}~~l>k \cr
& (W_{2p})_{lk} |\Psi_m^0>=|\Psi_m> ~~~~~~    {\rm if}~~l\le k
&(5.7) \cr}
$$
Therefore in the thermodynamic limit the Laughlin ground state is annihilated
by the infinitely many generators $W_{2p}$.
{}From the above construction it is obvious that on the
space of states $|\Psi_m>$, the
operators $W_{2p}[\xi]$ satisfy the same algebra as the $\rho[\xi]$
operators, namely the $W_{\infty}$ algebra
$$[W_{2p}[\xi _1], W_{2p}[\xi _2]]|\Psi _m> = W_{2p}[\{\!\!\{\xi _1 ,
\xi _2 \}\!\!\}]
|\Psi _m> \eqno(5.8)
$$
where $\{\!\!\{ \}\!\!\}$ is the Moyal bracket defined in (2.16).
Each $W_{2p}$, $p=0,1,...$
provides a representation for $W_{\infty}$ algebra.

We would further like to obtain a
second quantized expression for these operators in terms of fermionic creation
and annihilation
operators. Let us explicitly write down the transformation of a state in
${\cal{H}}_m$ under the action of $W_{2p}[\xi]$.
$$ \eqalignno{ & \delta ^{2p}_{\xi}|\Psi _m>  \equiv W_{2p}[\xi] |\Psi _m> =
U_{2p} ~\rho [\xi] ~\tilde {U} _{2p} |\Psi _m> & (5.9) \cr
& = \int d^2 z_1...d^2 z_N e^{-\sum |z_i|^2}
\prod_{i<j}(\bar{z}_i-\bar{z}_j)^{2p} \sum_{i} \ddag
\xi(\partial_{\bar{z}_i}, \bar{z}_i )\ddag \prod_{k<l}
(\bar{z}_k-\bar{z}_l)^{-2p} F(\bar{z}_1...\bar{z}_N) |z_1...z_N>
\cr}
$$
where $F(\bar{z}_1...\bar{z}_N)= \prod_{i<j}
(\bar{z}_i-\bar{z}_j)^{2p+1} P(\bar{z}_1...\bar{z}_N)$.

The second quantized expression for $W_{2p}[\xi]$ would
satisfy
$$ \delta _{\xi} ^{2p} |\Psi _m> = \int d^2 z_1...d^2 z_N
e^{-
\sum|z_i|^2} F(\bar{z}_1...\bar{z}_N) {1 \over \sqrt{N!}} \sum
_{i=1} ^{N} \psi^{\dag}(z_1)\psi^{\dag}(z_2)...[W_{2p},
\psi^{\dag}(z_i)]...\psi^{\dag}(z_N)|0> \eqno(5.10) $$
where $W_{2p}[\xi]|0>=0$.

Using eqs.(2.8) and (4.10) one finds that a candidate for the second quantized
operator $W_{2p}[\xi]$ is
$$W_{2p}[\xi]= \int d^2 z e^{-|z|^2} \psi^{\dag} (z) e^{2p \alpha
(\bar{z})} \ddag \xi ( \partial _{\bar{z}}, \bar{z}) \ddag e^{-2p \alpha
(\bar{z})} \psi(\bar{z}) \eqno(5.11)
 $$
where $\alpha (\zbar)$ is the operator defined in (4.9) and $e^{\alpha
(\zbar)}$ is the vortex operator. One can directly verify that
$W_{2p}[\xi]$ in (5.11) satisfies the $W_{\infty}$ algebra on ${\cal{H}}_m$ and
therefore provides a representation for $W_{\infty}$. Equation (5.11)
can be further written as
$$
W_{2p}[\xi]= \int d^2 z e^{-|z|^2} \psi^{\dag} (z) \ddag \xi
(\partial _{\bar{z}}-2p \int d^2 z' {{\rho (z', \bar{z}')} \over
{(\bar{z}-\bar{z}')}}, \bar{z}) \ddag \psi(\bar{z}) \eqno(5.12)
$$
The term $ 2p \int d^2 z' {{\rho (z',\bar{z}')} \over
{(\bar{z}-\bar{z}')}}$ plays the role of a gauge potential and it is similar
to the one induced by the Chern-Simons interaction.

Each of the two different forms (5.11) and (5.12) for $W_{2p}$ stresses a
different physical interpretation. In (5.12) $W_{2p}[\xi]$ is expressed in
terms of the usual fermions with a complicated interaction, similar to a
Chern-Simons interaction, while in (5.11) one can think of $W_{2p}[\xi]$ as a
bilinear (like $\rho [\xi]$ in $\nu=1$ case) of free composite fermions, made
out of the usual fermions and $2p$ vortices.

Having identified the operators $W_{2p}$ which create the excited states, it
is interesting to find their spectrum. The Haldane potential $V_{int} ^m$ in
(4.8) obviously commutes with $W_{2p}$ on the space of the wavefunctions (3.2).
On the other hand the confining potential $V_c$
is not of the form $W_{2p}[\xi]$. However if $V_c$ is a harmonic oscillator
potential as in (2.18), one can
show that
$$V_c=\lambda [(W_{2p})_{11} + p N (N-1)] \eqno(5.13) $$
where $N$ is the fermion number operator,
$N=\int d^2 z e^{-|z|^2} \psi ^{\dag}(z) \psi (\bar{z})$, which commutes
with any $W_{2p}$. As a result in the space ${\cal{H}}_m$
$$[V_c ~ , ~ (W_{2p})_{l,l+k}]= \lambda ~ k ~ (W_{2p})_{l,l+k} \eqno(5.14)$$
Comparing this with (3.7) we conclude that the slope of the excitation spectrum
of the Laughlin type states (3.2) is identical with the one in the $\nu =1$
case.

\vskip .3 in
\noindent{\bf 6. Two-body Interactions and $X_{\infty}$ algebra}

As we showed in the previous section the existence of a \Winf algebra in the
case of the $\nu=1/m$ FQHE described by Laughlin wavefunctions has to do with
the specific form of these wavefunctions and the one-to-one mapping with the
$\nu =1$ states.
Although the Laughlin states are quite successful in describing
the $\nu =1/m$ FQHE, one would like to understand better the
underlying dynamics, especially the role of the Coulomb repulsive
interactions, and the spectrum that emerges from it.

As we discussed in section 2, in the absence of two-body interactions, the
naturally
emerging algebraic structure is the $W_{\infty}$ algebra in (2.15). It is
essentially the
algebra of bilinears $C^{\dag} C$ which provide the most general
unitary transformations (linear in the space of $C$'s) which preserve
the particle number and the lowest Landau level condition. The introduction
of general two-body interactions projected onto the LLL via the
operators $O_{nm}, \bar{O}_{nm}$ as in eqs. (4.3)-(4.5), suggests the
extension of $W_{\infty}$
algebra by including bilinears in $C$'s and $C^{\dag}$'s, which change the
particle number by two. We call this extended algebra, the $X_{\infty}$
algebra.

In order to see more clearly the algebraic structure, it is instructive to
recall the
situation in the case of a finite number $N$ of fermions.
It is well known that one
can construct a Clifford algebra from $C_n$ and $C^{\dag}_n\ \ (n= 1, 2,
\cdots , N)$.
$$
\g_n =C_n+C^{\dag}_n \, \ \ \ \ \g_{N+n} = -i(C_n-C^{\dag}_n )\eqno (6.1)
$$
$$
\{\g_i ,\g_j \}=2\d_{ij} \ \ \ \ i, j = 1,2,\cdots ,2N\eqno (6.2)
$$
$\S_{ij}={1\over{2i}}[\g_i , \g_j ]$ form an $O(2N)$ Lie algebra and all the
fermion
states form the basis of one  $2^N$ dimensional
spinor representation of $O(2N)$.$^{[15]}$
{}From (6.1) one sees that
the generators $\S_{ij}$ of $O(2N)$ are the bilinears of all $C_n$ and
$C^{\dag}_n\ \ (n= 1, 2, \cdots , N)$. On the other hand the generators
of $U(N)$ are $C^{\dag}_n C_m \ \ (n, m = 1, 2, \cdots , N)$,
only the particle-number-preserving bilinears.
As $N \to \infty$, $U(N) \to W_{\infty}$ and $O(2N)\to X_{\infty}$.
Any LLL state is a component of a spinor representation of $X_{\infty}$.

The most convenient (mathematically) basis to describe the $X_{\infty}$
algebra is in terms of bilinears $\y (\zbar )$ and $\y ^{\dag} (z)$
$$
U(z , \zbar ')=\y^{\dag}(z )\y (\zbar '),\ \ \ O (\zbar , \zbar ')=\y (\zbar)
\y
(\zbar ')\, \ \ \ \  \bar{O}(z , z ')=\y^{\dag}(z')\y^{\dag}(z)\eqno (6.3)
$$
The commutation relations are
$$
\eqalign{
[ U(z_1 ,\zbar_2 ),U(z' _1 ,\zbar ' _2 )] &=e^{\zbar_2 z' _1}U(z_1 ,\zbar '_2 )
 - e^{\zbar ' _2 z_1}U(z' _1 ,\zbar _2 )\cr
[ U(z_1 ,\zbar_2 ), O (\zbar ' _1 ,\zbar '_2 )] &=e^{\zbar '_1 z_1}O(\zbar '_2
,\zbar _2 ) -e^{\zbar '_2 z_1}O(\zbar '_1 ,\zbar _2 ) \cr
[U(z_1 ,\zbar_2 ),\bar {O} (z'_1, z'_2 )]  &=e^{\zbar _2 z '_2}\bar {O} (z'_1,
z_1 )- e^{\zbar _2 z '_1}\bar {O} (z'_2, z_1 )\cr
[O (\zbar  _1 ,\zbar _2 ), \bar {O} (z'_1, z'_2 ) ]&=e^{\zbar_1 z'_1 +\zbar_2
 z'_2}
-e^{\zbar_1 z'_2 +\zbar_2 z'_1}\cr &\ \ +e^{\zbar_1 z'_2} U(z'_1 ,\zbar_2)-
e^{\zbar_2 z'_2} U(z'_1 ,\zbar_1) +e^{\zbar_2 z'_1} U(z'_2 ,\zbar_1)-
e^{\zbar_1 z'_1} U(z'_2 ,\zbar_2) \cr
[O(\zbar _1, \zbar _2), O(\zbar' _1, \zbar' _2)] &= [\bar{O}(\zbar _1, \zbar
_2),
\bar{O}(\zbar' _1, \zbar' _2)]=0 \cr} \eqno (6.4)
$$

Since the Hamiltonian in the presence of an external potential and a two-body
interaction involves the operators $\rho _{kl}, O_{kl}, \bar{O}_{kl}$, it is
useful to write down the commutation relations (CR) of these objects, which
will provide a
different basis for the $X_{\infty}$ algebra. The CR of $\rho$'s is given in
(2.15)
and is the usual $W_{\infty}$ algebra. It is straightforward to derive the CR
between $\rho$ and $\bar{O}$. We find
$$
\eqalign{ [\rho_{kl}, \bar{O}_{mn}]= & (-{1\over \sqrt{2}})^{k+l-2} {{k!l!}
\over
{\sqrt{n!m!}}}  \sum _{\delta =0} ^l \sum _{\g =max(0,k-l-m+\d)} ^{min(k,n+\d)}
{{(-1)^{\g + \d} (\delta +n)! (l+m-\d)!} \over {\d ! (l-\d)! \g !(k- \g)!}}
\cr
& \times
{1 \over {\sqrt{(l+m-k-\d+\g )! (n+ \d -\g)!}}}
\cases {\bar{O}_{l+m-k+\g -\d ,n+ \d -\g} & if $l+m+n \ge k$ \cr 0 &if $l+m+n
\le k$\cr}
\cr} \eqno (6.5)$$
By taking the complex conjugate of the above expression we find
$$[\rho_{kl}, \bar{O}_{mn}]^{\dag}=-[\rho_{lk}, O_{mn}]   \eqno(6.6)$$
The CR between $O$ and $\bar{O}$ is more complicated. First we have to express
$U(z_1 ,\zbar_2 )$ in terms of the density operator. We find that
$$
U(z_1 ,\zbar_2 )\equiv \y^{\dag}(z_1 )\y (\zbar _2)= {1\over {4\p}}e^{\zbar_2
z_1} \int d\a d\b \rho (\a,\b) \exp [-i z_1({{\a-i\b} \over 2})] \exp[-i
\zbar _2 ({{\a+i\b} \over 2})] \eqno(6.7) $$
where $\rho (\a,\b)$ is the Fourier transform of the fermion density
$$ \rho (\a,\b)= \int d^2 z e^{-|z|^2} e^{i(\a Rez+\b Imz)} \y^{\dag}(z)
\y (\zbar)
\eqno(6.8) $$
Expanding now (6.7) in powers of $z_1$, $\zbar _2$ we find
$$ U(z_1 ,\zbar_2 )= {{e^{\zbar_2 z_1}} \over {4 \pi}} \sum _{k=0}^{\infty}
\sum
_{l=0}^{\infty} \tilde{\rho} _{kl} {{z_1^k \zbar _2 ^l} \over {k!l!}}
\eqno(6.9)$$
where
$$
\eqalign{ \tilde{\rho} _{kl} &= (-i)^{k+l} \int d\a d\b \rho(\a,\b)
({{\a-i\b} \over 2})^k ({{\a+i\b} \over 2})^l \cr
&= (2 \pi)^2 (\partial _z)^k (\partial _{\zbar})^l
\rho(z,\zbar) | _{Rez=Imz=0} \cr}
\eqno(6.10) $$
Given (6.9) and (6.10) we can easily find that
$$
\eqalign{ [\bar{O}_{n0}, O_{l0}]=- 2 \d _{ln} + {1\over \pi}
\sqrt{{{n!l!} \over 2^{n+l}}} & \sum_{k=0} ^{\mu _{nl}} {{2^{\mu_{nl}-k}} \over
{(\mu _{nl}-k)! (\D _{nl}+k)! k!}} \cr
& \times \cases {\tilde{\rho} _{k, \D_{nl} +k} &if $l \ge n$ \cr \tilde{\rho}
_{\D_{nl} +k,k} &if $n \ge l$ \cr}
\cr} \eqno(6.11) $$
where $\mu _{nl}= min(n,l)$, $M_{nl}=max(n,l)$ and $\D _{nl}=
max(n,l)-min(n,l)$.

In order to find the more general CR for arbitrary modes $\bar{O}$ and $O$ it
is useful to introduce the translation operators
$$
\eqalign{ A &= \int d^2 z e^{-|z|^2} \yd (z) z \y (\zbar) \cr
A^{\dag} & = \int d^2 z e^{-|z|^2} \yd (z) \zbar \y (\zbar) \cr}
\eqno(6.12) $$
with the following commutation relations
$$
\eqalign{
& [A, A^{\dag}] = \int d^2 z e^{-|z|^2} \yd (z) \y (\zbar) \cr
&[A, O_{nm}] = -\sqrt{2(m+1)}~ O_{n,m+1}~~~~~  [A^{\dag}, O_{nm}] =
-\sqrt{2 m} ~O_{n,m-1}  \cr
&[A, \bar{O}_{nm}]  = \sqrt{2 m} ~ \bar{O}_{n,m-1}~~~~~~~~~~~~~~~
[A^{\dag}, \bar{O}_{nm}]
 = \sqrt{2 (m+1)} ~ \bar{O}_{n,m+1}  \cr
&[A, \tilde{\rho}_{nm}]  =
-\tilde{\rho}_{n,m+1}~~~~~~~~~~~~~~~~~~~~~~[A^{\dag},
\tilde{\rho}_{nm}]= \tilde{\rho}_{n+1,m} \cr
} \eqno(6.13)
$$
Using (6.13) we can express $O_{nm}$ and $\bar{O} _{nm}$ in terms of
$O_{n0}$ and $\bar{O}_{n0}$ respectively
$$ \eqalign{ O_{nm} & = (-{1 \over {\sqrt{2}}})^m {1 \over \sqrt{m!}} \partial
_p ^m ~ [e^{pA} O_{n0} e^{-pA}]_{p=0} \cr
\bar{O}_{nm} & = ({1 \over {\sqrt{2}}})^m {1 \over \sqrt{m!}} \partial
_q ^m ~ [e^{q A^{\dag}} \bar{O}_{n0} e^{-q A^{\dag}}]_{q=0} \cr}
\eqno (6.14) $$
and further the general commutator of arbitrary $\bar{O}$ and $O$ modes
in terms of (6.11)
$$ [\bar{O}_{nm}, O_{ls}]= (-{1 \over \sqrt{2}})^s ({1 \over \sqrt{2}})^m
{1 \over \sqrt{s!m!}} \partial _p ^s \partial _q ^m ~ [e^{pA} e^{q A^{\dag}}
[\bar{O}_{n0}, O_{l0}] e^{-q A^{\dag}} e^{-pA} e^{-2pq}]_{p=q=0} \eqno (6.15)
$$
Inserting (6.11) in the above expression we find
$$
\eqalign{
& [ \bar{O} _{nm}, O_{ls} ]= -2 \d _{nl} \d _{sm} + {1 \over \pi}
\sqrt{{{m!s!n!l!} \over {2^{\D _{nl}+\D _{ms}}}}} \sum_{r=0} ^{\mu _{ms}}
\sum_{k=0} ^{\mu_{nl}} {1 \over {2^{r+k} r!k!}} \cr
& \times {1 \over {(\mu_{nl}-k)! (\mu_{ms}-r)! (\D_{nl}+k)! (\D_{ms}+r)!}}
\cases {
\tilde{\rho} _{\D_{nl}+\D_{ms}+k+r,k+r} &if $ n \ge l,~ m \ge s$ \cr
\tilde{\rho} _{\D_{nl}+k+r,\D_{ms}+k+r} &if $ n \ge l,~ s \ge m$ \cr
\tilde{\rho} _{\D_{ms}+k+r,\D_{nl}+k+r} &if $ l \ge n,~ m \ge s$ \cr
\tilde{\rho} _{k+r,\D_{nl}+\D_{ms}+k+r} &if $ l \ge n,~ s \ge m$ \cr}
\cr} \eqno (6.16) $$
Equations (2.15), (6.5), (6.6) and (6.16) express the $X_{\infty}$ algebra.

There are many interesting issues regarding this algebra.
First it is worth noticing the structure of the CR (6.16). It suggests
that in the
limit where the second term on the rhs of (6.16) is zero, $\bar{O} _{mn}$ and
$O_{mn}$ play the role of creation and annihilation operators respectively
and the Hamiltonian can be easily diagonalized in the resulting Fock space.

We mentioned before that the Laughlin wavefunctions are eigenstates
of the Haldane potential. We expect this to be the case as we tune in a weak
confining potential. In section 5 we showed that in the space of Laughlin
wavefunctions the spectrum generating algebra is isomorphic to \Winf . On the
other hand, following the analysis of section 6, there is a corresponding
$X_{\infty}$ algebra. It is interesting to understand the relation between
these algebraic structures.

\bigskip
\noindent{\bf 7. Discussions}

In this paper we have presented an algebraic treatment of the QHE.

In the case of the IQHE $\nu =1$, where the Coulomb interactions can be
neglected, such
an approach has revealed the emergence of the \Winf algebra, which plays the
role of a spectrum generating algebra and expresses the symmetry of the ground
state$^{[7,10]}$.

In the case of the FQHE $\nu =1/m$ as expressed by the Laughlin wavefunctions,
an isomorphic algebra emerges. This has to do with the specific form of the
Laughlin wavefunctions and their one-to-one mapping to the $\nu =1$ states. The
generators of this \Winf algebra are expressed in a second quantized language
and they are written in terms of fermion and vortex operators.
A byproduct of the isomorphism of the algebraic structures associated to
$\nu =1$ and $\nu
=1/m$ cases is the fact that in the presence of a harmonic oscillator
potential the
dispersion relation of the corresponding edge states is identical.

It is interesting to see if this infinite algebraic structure survives and if
so its specific representation once we consider generalized wavefunctions
describing $\nu \ne 1/m$ FQHE.

Although the phenomenology of the FQHE based on variational wavefunctions
successfully accounts for the observed plateaus of the Hall conductivity
at specific filling fractions,
the precise nature of the true ground state and the creation of a gap in the
presence of Coulomb interactions is not quite understood analytically. An
algebraic approach to this problem leads to a new extended infinite algebraic
structure, the $X_{\infty}$ algebra. In section 6 we mentioned some interesting
issues regarding this algebra. Its significance in analyzing the spectrum of
FQHE is under investigation.

\vskip .3in
\noindent{\bf Acknowledgements}

I thank Satoshi Iso and Bunji Sakita for their collaboration at an early stage
of this work. Especially, I am grateful to Bunji Sakita for many valuable
discussions and suggestions.

This work was supported in part by the U.S. Department of Energy under the
contract number DE-FG02-85ER40231. I thank Bunji Sakita for his hospitality
in City College CUNY where part of this work was done and for partial support
from his NSF grant, PHY 90-20495.

\vskip 0.4in

While this work was being typed, two papers appeared, ref.[16] and ref.[17],
with some partially overlapping results and ideas.

\vskip 0.3 in
\centerline{\bf References}

\item{[1]} R.E. Prange and S.M. Girvin, {\it ``The Quantum Hall Effect''},
Springer, New York, 1990.
\item{[2]} M. Stone, {\it ``Quantum Hall Effect''}, World Scientific, 1992.
\item{[3]} R.B. Laughlin in ref.[1]; {\it Phys. Rev. Lett.} {\bf 50} (1983)
1395.
\item{[4]} F.D.M. Haldane in ref. [1].
\item{[5]} F.D.M. Haldane, {\it Phys. Rev. Lett.} {\bf 51} (1983) 605; B.I.
Halperin, {\it Phys. Rev. Lett.} {\bf 52} (1984) 1583.
\item{[6]} J.K. Jain, {\it Phys. Rev. Lett.} {\bf 63} (1989) 199; {\it Phys.
Rev.} {\bf B40} (1989) 8079; {\it Phys. Rev.} {\bf B41} (1990) 8449.
\item{[7]} S. Iso, D. Karabali and B. Sakita, {\it Phys. Lett.} {\bf B296}
(1992) 143.
\item{[8]} M. Stone, {\it Phys. Rev.} {\bf B42} (1990) 8399.
\item{[9]} X.G. Wen, {\it Int. J. Mod. Phys.} {\bf B6} (1992) 1711 and
references therein.
\item{[10]} A. Capelli, C. Trugenberger and G. Zemba, {\it Nucl. Phys.}
{\bf B396} (1993) 465.
\item{[11]} N. Read, {\it Phys. Rev. Lett.} {\bf 62} (1989) 315.
\item{[12]} I. Bakas, {\it Phys. Lett.} {\bf
B228} (1989) 57; D.B. Fairlie, P. Fletcher and C.K. Zachos, {\it Phys. Lett.}
{\bf B218} (1989) 203; J. Hoppe and P. Schaller,{\it
Phys. Lett.} {\bf B237} (1990) 407; C.N. Pope, L.J. Romans and X. Shen {\it "A
Brief History of $W_{\infty}$''} in Strings 90, ed. Arnowitt et al (World
Scientific 1991) and references therein.
\item{[13]} S.M. Girvin and T. Jachs, {\it Phys. Rev.} {\bf B29} (1983) 5617;
C.
Itzykson, in {\it ``Quantum Field Theory and Qantum Statistics''}, A. Hilger,
Bristol,1986.
\item{[14]} F.D.M. Haldane, {\it Phys. Rev. Lett.} {\bf 51} (1983) 1395; S.A.
Trungman and S. Kivelson, {\it Phys. Rev.} {\bf B31} (1985) 5280.
\item{[15]} R.N. Mohapatra and B. Sakita, {\it Phys. Rev.} {\bf D21} (1980)
1062.
\item{[16]} M. Marsili, SISSA preprint, September 1993.
\item{[17]} M. Flohr and R. Varnhagen, BONN-HE-93-29 preprint, September 1993.
\end